\documentclass[aps,preprint,amsmath,amssymb,amsfonts,nofootinbib]{revtex4-1}
\usepackage{epsfig}
\usepackage{graphicx}
\usepackage{dcolumn}
\usepackage{bm}
\usepackage{amsthm}
\usepackage{amsmath}
\usepackage{color}

\newcommand{\beq}{\begin{equation}}
\newcommand{\eeq}{\end{equation}}
\newcommand{\bea}{\begin{eqnarray}}
\newcommand{\eea}{\end{eqnarray}}

\begin{document}

\title{Spontaneously Broken Asymptotic Symmetries and an Effective Action for Horizon Dynamics}
\author{Christopher Eling}
\email{cteling@gmail.com}
\affiliation{Rudolf Peierls Centre for Theoretical Physics, University of Oxford, 1 Keble Road, Oxford OX1 3NP, UK}

\begin{abstract}

Asymptotic spacetime symmetries have been conjectured to play an important role in quantum gravity. In this paper we study the breaking of asymptotic symmetries associated with a null horizon boundary. In two-dimensions, these symmetries are reparametrizations of the time parameter on the horizon. We show how this horizon reparametrization symmetry is explicitly and spontaneously broken in dilaton gravity and construct an effective action for these pseudo-Goldstone modes using the on-shell gravitational action for a null boundary. The variation of this action yields the horizon constraint equation. This action is invariant under a 2 parameter subgroup of $SL(2)$ transformations, whose Noether charges we interpret via the membrane paradigm. We place these results in the context of recent work on the near $AdS_2$/ near $CFT_1$ correspondence. In this setting the horizon action characterizes the infrared regime near the horizon and has a hydrodynamical sigma model form. We also discuss our construction in General Relativity. In the three-dimensional case there is a natural generalization of our results. However, in higher dimensions, the variation of the effective action only yields the Raychaudhuri equation for small perturbations of the horizon.

\end{abstract}

\maketitle

\section{Introduction}

Recently there has been renewed interest in the asymptotic symmetries associated with black hole horizons. In this case one considers the subset of diffeomorphisms that preserve the boundary conditions associated with the existence of a null surface in the metric, for early works see \cite{Carlip:1999cy, Hotta:2000gx, Koga:2001vq}. These can be thought of as symmetries of the horizon system, potentially responsible for the universality of the Bekenstein-Hawking entropy via, for example, the central charge of a Virasoro algebra.  Recently it was shown that this analysis generically leads to a set of infinite dimensional symmetries that is closely analogous \cite{Donnay:2015abr} to the Bondi-Metzner-Sachs (BMS) symmetries associated with null infinity in asymptotically flat spacetimes \cite{Bondi}. The horizon BMS symmetries are associated with ``soft hair" on black holes, which potentially can help to resolve the black hole information loss paradox \cite{Hawking:2016msc}. For related works see for example \cite{Penna:2015gza, Blau:2015nee, Averin:2016ybl,Compere:2016hzt,Afshar:2016wfy,Averin:2016hhm,Hotta:2016qtv,Afshar:2016uax,Mirbabayi:2016axw}.

The near-horizon geometry of a black hole in $D$ dimensions can be expressed in null Gaussian coordinates $(v,x^i,r)$ with $i = 1..D-2$
\begin{align}
ds^2 = \left( -2 \kappa(v,x^i) r + O(r^2) \right) dv^2 + \left(4 \Omega_i(v,x^i) r + O(r^2) \right) dx^i dv + 2 dv dr + \gamma_{ij}(v,x^i,r) dx^i dx^j. \label{nullgaussian}
\end{align}
The horizon is located at $r=0$. The quantity $\kappa$ is generally a measure of the non-affinity of the horizon's null geodesic generators
\begin{align}
\ell^B \nabla_B \ell^A = \kappa \ell^A. \label{geodesic}
\end{align}
$\Omega_a$ is an extrinsic curvature one form
\begin{align}
\Omega_i = k_B \nabla_A \ell^B,
\end{align}
where $k_A$ is a null vector such that $k_A \ell^A = 1$.

One can consider diffeomorphisms that satisfy the following gauge fixing conditions
\begin{align}
{\cal L}_\xi  g_{rr} = 0, ~~ {\cal L}_\xi  g_{vr} = 0,\label{cond1}
\end{align}
along with horizon preserving conditions
\begin{align}
{\cal L}_\xi  g_{vv} = 0 + O(r), ~~ {\cal L}_\xi g_{vi} = 0 + O(r). \label{cond2}
\end{align}
The vector field generating this class of (infinitesimal) diffeomorphisms has the generic form
\begin{align}
\xi^A \partial_A = \epsilon(v,x^i) \partial_v + \left(R^i(x^i) - r \gamma^{ij} \partial_j \epsilon(v,x^i)  \right) \partial_i  - \left(r \partial_v \epsilon(t,x^i) - r^2 \Omega_i \partial^i \epsilon(v,x^i)  \right)\partial_r + \cdots \ , \label{xisoln}
\end{align}
where $\epsilon$ and $R^i$ are arbitrary functions. The function $\epsilon$ generates ``supertranslations", while $R^i$ generates ``superrotations"; these names are chosen in analogy with BMS transformations at null infinity.

Here we will concentrate on the behavior of the horizon supertranslations. Unlike asymptotic BMS supertranslations at null infinity, here $\epsilon$ can be a function both of time and space. If we consider the simplest case of a two dimensional black hole geometry
($D=2$) then no superrotations are possible, but there is a remaining horizon ``supertranslation" freedom
\begin{align}
v \rightarrow v + \epsilon(v),
\end{align}
amounting to a time reparametrization freedom on the horizon. The same time reparametrization freedom appears as the asymptotic symmetry of $AdS_2$ spacetimes,  see e.g. \cite{Hotta:1998iq, Maldacena:2016upp}. In the AdS/CFT correspondence, the time reparametrizations amount to one-dimensional conformal transformations of the boundary metric $ds^2 = -dt^2$.  

Recent work has shown that the proper formulation of holography is in terms of a near $AdS_2$/near $CFT_1$ duality \cite{Almheiri:2014cka, Jensen:2016pah, Maldacena:2016upp, Engelsoy:2016xyb}. On the gravity side, one works with dilaton gravity \cite{Grumiller:2002nm}, allowing for a non-trivial dilaton field in the bulk $AdS_2$.  The presence of the dilaton  breaks the time parametrization symmetry explicitly. There is also a spontaneous breaking by the choice of the $AdS_2$ vacuum.  Therefore the system is characterized by reparametrization mode pseudo-Goldstones. Maldacena, Stanford, and Yang showed the equations of motion for dilaton gravity imply a relationship between the (renormalized) boundary value of the dilaton $\phi_r$ and the reparametization mode $t(u)$ (i.e. $t \rightarrow t(u)$). This equation can be derived from the gravitational boundary action, which is an effective Schwarzian action invariant under $SL(2)$ transformations \cite{Maldacena:2016upp}. Interestingly, \cite{Jensen:2016pah} showed that this action can be expressed in a hydrodynamical form following \cite{Haehl:2015pja}, where the basic variables are maps between a reference manifold and the physical spacetime (e.g. the mapping between the Lagrangian and Eulerian descriptions of a fluid). Indeed, the reparametrization mode is a mapping between two times.

In this paper we study the breaking of the horizon reparametrization symmetry and show that it governs a type of out of equilibrium dynamics of the horizon membrane system.  In two-dimensional gravity we show that the pattern of reparametrization symmetry breaking at the horizon is analogous to the behavior at asymptotic infinity in $AdS_2$. The horizon constraint equation for dilaton gravity yields a relation between the reparametrization mode $\alpha(v)$ and the horizon value of the dilaton field $\phi_H$. This equation can also be derived from an effective action, which is the on-shell gravitational action for a null boundary. This action has symmetries which turn out to be associated with time reparametrizations that preserve the non-affinity $\kappa$ of the horizon. The associated conserved Noether charges can be interpreted via the membrane paradigm, following \cite{Eling:2016xlx}. We can also express this action in a hydrodynamical sigma model form.

In the case of the $AdS_2$ black hole, there are in principle two boundaries, one at infinity and the other at the horizon. In equilibrium, the horizon terms are topological, but for perturbations around equilibrium they describe the behavior of the horizon membrane and can be thought of as capturing the infrared degrees of freedom in a dual field theory. The Noether charges associated with the infrared action turn out to be two of the three charges associated with the $SL(2)$ symmetry of the Schwarzian action.

Some of these results generalize when we consider higher dimensional gravity, where the horizon constraint equation is the Raychaudhuri equation. In $D=3$ the Raychaudhuri equation simplifies and effectively one just has to map $\phi_H$ into the determinant of the horizon metric, $\sqrt{\gamma}$. Therefore a sector of the horizon dynamics is controlled by the generalized mode $\alpha(v,x)$ via the same type of hydrodynamical action. For horizons in $D>3$, the story is more complicated. Here the Raychaudhuri equation has shear squared terms, which act as a dissipation due to gravitational waves crossing the horizon. Even if we absorb these into a generalized matter-energy flux across the horizon, our action still only describes linearized perturbations of the Raychaudhuri equation.

The organization of this paper is as follows. In Section II we review the gravitational construction of the Schwarzian action in $AdS_2$ via symmetry breaking. In Section III we follow a similar logic to derive the effective action at any horizon in two-dimensional gravity. We examine the symmetries of this action, interpretation of the Noether charges, and how one can consider contributions from matter fields.  In Section IV we discuss the role of this action in the recently proposed near $AdS_2$/near $CFT_1$ correspondence. In Section V and VI we generalize our results to three and higher dimensional gravity. We conclude with a discussion of future directions.

\section{Reparametrizations and their breaking in nearly $AdS_2$}
\label{AdSBoundary}

Here we briefly review the time reparametrization symmetry and its breaking in nearly $AdS_2$ spacetime, following \cite{Maldacena:2016upp}. Consider first the (Euclidean) $AdS_2$ vacuum in Poincare coordinates
\begin{align}
ds^2 = \frac{dt^2+dz^2}{z^2}  \label{Poincare}
\end{align}
and a curve defined by $(t(u), z(u))$ where $u$ is the internal time parameter. The curve defines a timelike boundary in the spacetime. As we approach the AdS boundary, the proper time on the curve will diverge. Therefore we introduce a cutoff $\ell_c$ such that
\begin{align}
\frac{1}{\ell_c} = \sqrt{\frac{t'^2+z'^2}{z^2}},
\end{align}
where primes indicate derivatives. From this equation, we deduce that for small $\ell_c$, $z = \ell_c t' + \cdots$. Thus, a given $t(u)$ specifies the boundary cut-off trajectory.

As an asymptotic symmetry, the one-dimensional conformal symmetry/time reparametrization maps one $AdS_2$ spacetime into another and therefore one boundary curve into another via $t(u) \rightarrow t(u) + \epsilon(u)$. The choice of the $AdS_2$ vacuum state spontaneously breaks the infinite dimensional symmetry. In this case it is broken down to the group of $SL(2,R)$ symmetries that preserve the vacuum state. In particular, under the global $SL(2,R)$ transformation
\begin{align}
t \rightarrow \frac{at + b}{ct+d}, ~~ ad-bc=1,
\end{align}
the boundary cut-off shape is unchanged.

In two-dimensions Einstein gravity is trivial. Since the Ricci scalar is a total derivative, the Einstein-Hilbert action is topological and there is no dynamics. On the other hand, one can define a gravitational theory with a scalar degree of freedom, the dilaton. The dilaton arises from the dimensional reduction of a higher dimensional gravitational theory to two-dimensions.  A simple early model of dilaton gravity is the Jackiw-Teitelboim (JT) theory \cite{JT},
\begin{align}
I_{JT} = \frac{1}{16\pi G_N} \left[ \int d^2x \sqrt{g} \phi ~ \left( R+ 2 \right) + 2 \int dt \phi_b K \right], \label{JTaction}
\end{align}
where in the additional Gibbons-Hawking term $\phi_b$ is the boundary value and $K$ is the extrinsic curvature of the boundary. The equations of motion imply that the solutions are $AdS_2$ with the non-trivial dilaton profile
\begin{align}
\phi = \frac{a_1 + a_2 t + a_3(t^2 + z^2)}{z}. \label{AdSdilaton}
\end{align}
Note that if we consider the diffeomorphisms $\phi \rightarrow \phi + \xi^A \partial_A \phi$ that preserve the asymptotic form of the dilaton ($z^{-1}$ divergent piece) then the arbitrary time reparametrization is broken explicitly down to $SL(2)$.

The boundary value $\phi_b$ can be expressed in terms of a renormalized $\phi_r(u)$ on the cutoff boundary via $\phi_b = \phi_r(u)/\ell_c$. Using (\ref{AdSdilaton}), one finds
\begin{align}
\frac{a_1 + a_2 t(u) + a_3 t(u)^2}{t'(u)} = \phi_r(u) \label{dilatonfixing}
\end{align}
relating the dilaton coupling to the reparametrization mode. One can derive (\ref{dilatonfixing}) from the variation of the on-shell gravitational action, which is the Gibbons-Hawking term in (\ref{JTaction}). Note that one has imposed the dilaton equation of motion, but not the metric field equation. Evaluating the extrinsic curvature of the boundary curve, one finds
\begin{align}
I = -\frac{1}{8\pi G_N} \int du \phi_r(u) \{t,u\},
\end{align}
where $\{t,u\}$ is the Schwarzian derivative
\begin{align}
\{t,u\}  = \frac{t'''(u)}{t'(u)}- \frac{3}{2}\frac{t''(u)^2}{t'(u)^2}.  \label{Schwarzian}
\end{align}
The action is invariant under global $SL(2,R)$ transformations and one can study the associated conserved charges. The Schwarzian derivative also describes the low energy (strongly coupled), large $N$ regime of the SYK model, which is a one-dimensional quantum mechanical theory with $2N$ Majorana fermions (see, for example \cite{Kitaev, Maldacena:2016hyu}).

To describe the $AdS_2$ black hole case, one can re-analyze the system with metric given by, for example
\begin{align}
ds^2 = -\sinh^2 \rho d\bar{\tau}^2 + d\rho^2,
\end{align}
or simply redefine $t(u) = \tanh(\pi \bar{\tau}(u)/\beta_0)$, which is the conformal transformation from the vacuum to a state at finite temperature $\beta_0^{-1}$. The resulting effective action is
\begin{align}
I = -\frac{1}{8\pi G_N} \int du \phi_r(u) \left( \{\bar{\tau},u\} + \frac{2\pi}{\beta_0} \bar{\tau}'^2 \right).   \label{boundaryaction}
\end{align}

Jensen showed that this action can be expressed in hydrodynamical sigma model form \cite{Jensen:2016pah}. We think of $u$ as the time coordinate on the physical spacetime, the boundary.  $\bar{\tau}$ labels time on a reference manifold $M$. The metric on $M$ is
\begin{align}
h = -u'(\bar{\tau})^2 d\bar{\tau}^2.
\end{align}
On the reference manifold one works with a fixed vector field $\beta^a$ such that the temperature and velocity are defined as
\begin{align}
T = \frac{1}{\sqrt{-h_{ab} \beta^{a} \beta^{b}}}, ~~ u^{a} = \frac{\beta^{a}}{\sqrt{-h_{ab} \beta^{a} \beta^{b}}}.
\end{align}
In one-dimension we have $\beta^{\bar{\tau}} = \beta_0 = T_0^{-1}$, yielding
\begin{align}
T = \frac{T_0}{\sqrt{-h}}, ~~ u^\alpha = \frac{1}{\sqrt{-h}}.
\end{align}
In terms of these variables the action (\ref{boundaryaction}) has the general hydrodynamical form
\begin{align}
I_{eff} = \frac{1}{8\pi G_N} \int d\bar{\tau} \sqrt{-h} \phi_r \left(\frac{3}{2} \frac{\dot{T}^2}{T^2} - \frac{\ddot{T}}{T} + 2 \pi^2 T^2 \right). \label{hydroboundary}
\end{align}
In the case where $\phi_r$ is a constant, we can eliminate total derivatives to find the form given in \cite{Jensen:2016pah}.

\section{Horizon reparametrizations and effective action}

We now describe how an analogous pattern of symmetry breaking exists in the two-dimensional horizon system. In the horizon setting, the remaining ``supertranslation" symmetry can be thought of as the freedom to reparametrize the time along the null geodesics of the horizon surface.
We will allow for a generic dilaton theory
\begin{align}
I_{dil} = \frac{1}{16\pi G_N} \left[ \int d^2x \sqrt{g} \left( \phi R+ V(\phi) \right) + 2 \int du \phi_b K \right], \label{dilatonaction}
\end{align}
where the potential $V(\phi)$ is arbitrary ($V(\phi)=2\phi$ gives the JT theory). A black hole solution has the near-horizon metric (in null Gaussian coordinates)
\begin{align}
ds^2 = (-2 \kappa r + \cdots )dv^2 + 2 dv dr.
\end{align}
At this stage, the asymptotic behavior of this solution at infinity is arbitrary, we are focusing strictly on the near-horizon physics. The transformations $v \rightarrow \alpha(v)$ map the solution into a different black hole with different $\kappa$. The metric field equation is
\begin{align}
E_{AB} = \nabla_A \nabla_B \phi - g_{AB} \Box \phi +  g_{AB} V(\phi) = 0.  \label{metricfieldeqn}
\end{align}
Contracting with two null normals $\ell^A$ and evaluating on the horizon yields the horizon constraint equation
\begin{align}
\ell^A \ell^B \nabla_A \nabla_B \phi_H = \ell^A \nabla_A (\ell^B \nabla_B \phi_H) - \kappa \ell^B \nabla_B \phi_H = 0. \label{horizonconstraint}
\end{align}
The solution to this equation in a general time parametrization is
\begin{align}
\phi_H(v) = c_0 + c_1 \int^{\alpha(v)} dt' e^{\int^{v''} \kappa(v'') dv'' }.
\end{align}
For simplicity, we will typically consider the case where $\kappa$ is a constant, which yields
\begin{align}
\phi_H(v) = c_0 + \frac{c_1}{\kappa} e^{\kappa \alpha(v)}. \label{dilatonfixing2}
\end{align}
This gives us a relation between the reparametrization mode $\alpha(v)$ and the boundary value $\phi_H$ in analogy with (\ref{dilatonfixing}) at the $AdS_2$ boundary.

The presence of the non-trivial dilaton on the horizon breaks the infinite dimensional reparametrization symmetry. We should require that diffeomorphisms $\phi \rightarrow \phi + \xi^A \partial_A \phi$ also preserve asymptotic form of the dilaton solution,
\begin{align}
\phi(v,r) = c_0 + \frac{c_1}{\kappa} e^{\kappa \alpha} + O(r).
\end{align}
We find that
\begin{align}
\xi^t =  \frac{G}{\alpha'} + \frac{F}{\alpha'} e^{-\kappa \alpha} \label{alphatransform}
\end{align}
preserves this form, for constants $G$ and $F$, mapping into another solution with different constants $c_0$ and $c_1$.

Note that just as in the $AdS_2$ boundary case there is also a spontaneous breaking of the reparametrization symmetry, coming from our choice of state with a particular $\kappa$. Under the infinitesimal reparametrization
\begin{align}
\kappa \rightarrow \kappa + \epsilon \partial_v \kappa + \partial^2_v \epsilon + \kappa \partial_v \epsilon  \ . \label{kappachange}
\end{align}
Neglecting the time dependence in $\kappa$, we that the condition for $\kappa$ to be preserved is \cite{Eling:2016xlx}
\begin{align}
\epsilon(v) = G + F e^{-\kappa v},
\end{align}
which agrees with (\ref{alphatransform}). For finite transformation one can find the change in the surface gravity by reparametrizing the time $\bar{v} = \alpha(v)$ in the null geodesic equation (\ref{geodesic}). The result is
\begin{align}
\kappa  =  \bar{\kappa} \alpha' + \frac{\alpha''}{\alpha'} \label{finitekappa}
\end{align}
Setting $\kappa = \bar{\kappa}$ in (\ref{finitekappa}),  yields the general transformation
\begin{align}
\alpha(v) = \frac{\ln (k_1 e^{\kappa v} + k_0)}{\kappa}. \label{finitetrans}
\end{align}

Interestingly, it is also possible to derive the condition (\ref{dilatonfixing2}) from an effective action, analogous to that of the $AdS_2$ boundary. We can construct the effective action for the reparametrization modes in the following way. For a timelike boundary, one must supplement the gravitational  action with the Gibbons-Hawking term in (\ref{dilatonaction}). The on-shell gravitational action reduces to just this boundary term, which should correspond to the effective action for the Goldstone modes. For a null surface, one can evaluate the dilaton boundary term in null geodesic coordinates and then take the null $r=0$ limit. The result is
\begin{align}
I_{null}  = \frac{1}{8\pi G_N} \int dv \phi_H \kappa,
\end{align}
which is consistent with earlier results for null boundaries \cite{Parattu:2015gga}. We take $\kappa$ to have the form $\kappa = \frac{\alpha''}{\alpha'} + \alpha' \kappa_0$, with $\kappa_0$ a
fixed background surface gravity associated with an equilibrium state. We propose that the effective action has the form
\begin{align}
I_{eff} =   \frac{1}{8\pi G_N}  \int dv \phi_H \left(\frac{\alpha''}{\alpha'} + \alpha' \kappa_0 \right) , \label{action}
\end{align}
where $\phi_H$ acts like an external coupling. Variation of the field $\alpha(v)$ produces the equation of motion
\begin{align}
\left(\frac{\phi'_H}{\alpha'}\right)' - \kappa_0 \phi'_H = 0.
\end{align}
We can re-express this in the form
\begin{align}
\phi''_H - \left(\frac{\alpha''}{\alpha'} + \kappa_0 \alpha' \right) \phi'_H = 0,
\end{align}
which is exactly the general form of the horizon constraint equation in  (\ref{horizonconstraint}). Therefore the one-dimensional gravity action captures the near-horizon physics. Note that if $\phi_H$ is a constant, then the action is a total derivative and the dynamics are trivial. Therefore (\ref{action}) describes a type of non-equilibrium dynamics of the horizon, characterized by Goldstones $v \rightarrow v + \epsilon(v)$.

We can also consider the special case where the horizon is extremal and $\kappa_0 = 0$. Here the transformation preserving extremality is $\epsilon(\lambda) = G + F \lambda$, which is an affine transformation, with $\lambda$ the affine parameter. The effective action 
takes the form
\begin{align}
I^{ext}_{eff} =   \frac{1}{8\pi G_N}  \int d\lambda \phi_H \left(\frac{\alpha''}{\alpha'} \right). \label{extremalaction}
\end{align}

Finally, we note in passing that one may wonder about the generic case of a time dependent $\kappa_0$. Here the form of the effective action is less clear, but we conjecture the action is
\begin{align}
I_{eff} =   \frac{1}{8\pi G_N}  \int dv \phi_H \left(\frac{\alpha''}{\alpha'} + \alpha' \kappa_0 + \alpha \kappa'_0 \right), \label{actiontimedep}
\end{align}
which again leads to the horizon constraint equation when we vary $\alpha$.

\subsection{Symmetries and Noether charges}

We now consider the Noether charges associated with the horizon action (\ref{action}).  As expected, it is invariant under the infinitesimal global symmetry in (\ref{alphatransform}), i.e $\alpha \rightarrow \alpha + G + F e^{-\kappa_0 \alpha}$. Computing the conserved Noether charges associated with the symmetry yields the two parameter family
\begin{align}
Q = \frac{1}{8\pi G_N} \left[G \left(\kappa_0 \phi_H - \frac{\phi'_H}{\alpha'} \right) - F \phi'_H \frac{e^{-\kappa_0 \alpha}}{\alpha'} \right]  \label{Qir}
\end{align}
Note that the first charge proportional to $G$ is conserved trivially by horizon constraint equation, which can be expressed as $\frac{dQ_G}{dt} = 0$. If we consider small perturbations in $\alpha$, we find
\begin{align}
Q_G =&  \frac{1}{8\pi G_N} \left(\kappa_0 \phi_H - \phi'_H \right) \\
Q_F =& - \frac{1}{8\pi G_N} \phi'_H e^{-\kappa_0 v}.
\end{align}
In the equilibrium case where $\phi_H$ is a constant, we find
\begin{align}
Q^{(0)}_G= T_0 s^{(0)},
\end{align}
where $T_0 = \frac{\kappa_0}{2\pi}$ and entropy $s^{(0)} = \frac{\phi^{(0)}_H}{4G_N}$. This reproduces the Wald entropy formula in the stationary case, which tells us that entropy is proportional to a time translation Noether charge \cite{Iyer:1994ys}.

The action in the extremal case (\ref{extremalaction}) is invariant under the global affine transformation $\alpha \rightarrow \alpha + G + F \alpha$. Here the two corresponding Noether charges are
\begin{align}
Q^{ext}_G=  -\frac{\phi'_H}{\alpha'} \\
Q^{ext}_F = \phi_H - \frac{\phi'_H}{\alpha'} \alpha.
\end{align}

To investigate further the physical properties of the conserved charges we follow \cite{Eling:2016xlx} and consider membrane paradigm picture, where the expectation value of the Brown-York stress tensor is thought of as the stress tensor of the horizon field theory system.  The Brown-York stress tensor is defined as the canonical momentum with respect to the induced metric on the hypersurface. However, in one-dimension, there is only one component, the energy. To compute the energy for dilaton gravity, we consider the action (\ref{dilatonaction}).
The 1+1 dimensional decomposition of the Ricci scalar yields
\begin{align}
R = - \nabla_A (n^A \nabla_C n^C) + \nabla_C (n^A \nabla_A n^C)
\end{align}
where $n^A$ is the normal to the timelike slice. Integrating by parts, the action can be expressed as
\begin{align}
S = \frac{1}{16 \pi G_N} \int N \sqrt{h} d^2 x \left(n^A \nabla_A \phi) (\nabla_C n^C) - \nabla_C \phi (n^A \nabla_A n^C) + V(\phi) \right)
\end{align}
To find the canonical momentum we are interested in terms containing radial derivatives of the induced one-dimensional metric $h_{tt}$. The result is
\begin{align}
 E = -\frac{1}{8 \pi G_N} n^A \nabla_A \phi.   \label{Etimelike}
\end{align}
In the horizon limit the membrane energy is
\begin{align}
E_{memb} =  -\frac{1}{8 \pi G_N} \ell^A \nabla_A \phi_H.
\end{align}
Therefore we can re-express the conserved Noether charges as
\begin{align}
Q = G (T_0 s + E_{memb}) + F E_{memb}~ e^{-\kappa_0 \alpha},
\end{align}
where we take $s \sim \phi_H(v)$. The second conserved charge (term proportional to $F$) is associated with the energy of the membrane system. If we take the time derivative of the second charge, we find the expected $dQ/dv=0$ by virtue of the constraint equation (\ref{horizonconstraint}) which can be written as $\ell^A \nabla_A E_{memb} - \kappa_0 E_{memb} = 0$. Of course, for our solution (\ref{dilatonfixing2}), $E_{memb} = c_1 e^{\kappa_0 \alpha}$. 

In the extremal case we find 
\begin{align}
Q^{ext} = G E_{memb} + F (s + E_{memb} \alpha)
\end{align}
Here horizon constraint equation implies $dE_{memb}/d\lambda$, so that the membrane energy is a constant.

\subsection{Adding Matter}

Now suppose that we add matter-energy to the horizon system. In this case the horizon constraint equation is modified to
\begin{align}
\ell^A \nabla_A E_{memb} - \kappa E_{memb}  = 2 T_{AB} \ell^A \ell^B  \label{mattereqn}
\end{align}
where $T_{AB}$ the stress-energy tensor for the matter fields. The flux of matter-energy leads to non-conservation of the membrane energy.  As an example, we first consider the case where a light-like shell of mass $M$ falls across the horizon. Here the stress tensor has the form
\begin{align}
T_{\bar{v} \bar{v}} = M \delta(\bar{v}-v_0)
\end{align}
where $\ell^A \nabla_A = \partial_{\bar{v}}$. Solving (\ref{mattereqn}) we find
\begin{align}
E_{memb} = c_1 e^{\kappa \bar{v}} + M e^{\kappa (\bar{v}-v_0)}(1 - \theta(v_0-\bar{v}))
\end{align}
Thus we see that the mass of the shell has been incorporated into the conserved energy of the membrane system. In affine parametrization the result is somewhat cleaner: here $E_{memb} = c_1 + M(1-\theta(\lambda_0-\lambda))$. After the passage of the shell
there is a shift in the membrane energy of magnitude $M$.

We can understand this effect more generally by considering the stress tensor for scalar field(s), $\psi$. In this case the contribution to (\ref{mattereqn}) is of the form
\begin{align}
T_{AB} \ell^A \ell^B = \ell^A \ell^B \nabla_A \nabla_B \psi = \ell^A \nabla_A (\ell^B \nabla_B \psi) - \kappa \ell^A \nabla_A \psi.
\end{align}
Any terms proportional to $g_{AB}$ in the stress tensor vanish due to the contraction with null vectors. Therefore we can write an effective action for the contribution of scalar matter as
\begin{align}
I_{matt} = \int dt \psi_H \kappa = \int dv \psi_H  \left(\frac{\alpha''}{\alpha'} + \kappa_0 \alpha' \right).
\end{align}
The variation of $I_{matt}$ plus our earlier gravitational action (\ref{action}) with respect to the reparametrization mode $\alpha$ yields the full horizon constraint equation (\ref{mattereqn}). Therefore the matter fields also carry the two Noether charges we discussed earlier. In the total system
\begin{align}
Q_{tot} = Q_{grav} + Q_{matt}
\end{align}
remains conserved.

Another thing we can do is take our effective action (\ref{action}) and couple it to a quantum effective action that arises from integrating out matter fields on the two-dimensional spacetime background. The quantum effective action encodes information about the renormalized stress tensor operator. In this way we can describe a semi-classical backreaction process (such as evaporation).  If we consider conformal matter, then the effective action is controlled by the conformal anomaly, via the Polyakov action
\begin{align}
I_p = \int d^2 x \sqrt{g} R \Box^{-1} R
\end{align}
We can write this in a local form via the auxiliary scalar field $\chi$
\begin{align}
I_p = \int d^2 x \sqrt{g} \left(\partial_\mu \chi \partial^\mu \chi + \chi R \right) + \int dt \chi \kappa,
\end{align}
where we have included a Gibbons-Hawking like term in the presence of the null boundary. This boundary term has the form of our matter effective action. Inserting $\kappa = \frac{\alpha''}{\alpha'} + \alpha' \kappa_0$ and integrating over $\chi$, one can in principle find a non-local action in terms of the scalar Green's function.

\section{$AdS_2$ Black Hole case}

In the previous analysis we have worked solely at the horizon and used no information about the global behavior of the solution. However,  one can ask about the about the interpretation of (\ref{action}) in terms of the near $AdS_2$/ near $CFT_1$ correspondence when we have an $AdS_2$ black hole solution. We start with the metric in Eddington-Finkelstein coordinates
\begin{align}
ds^2 = -(\rho^2-\rho_0^2) d\tau^2 + 2 d\tau d\rho,
\end{align}
where $\rho_0$ is a constant. If we then consider the diffeomorphisms that preserve the AdS metric, which as we described in Section II amount to one-dimensional conformal transformations, we find
\begin{align}
ds^2 = -(\rho^2 - \rho_s^2) du^2 + 2 du d\rho .  \label{AdSBH}
\end{align}
Here $\rho_s^2(u) = -2 \{ \tau,u\} + \rho_0^2 \tau'^2$. The horizon radius a priori depends on time and can be found from the condition that the null normal $\ell_A = \partial_A (\rho-\rho_h(u))$ is null. The result is that $\kappa(t) = \rho_h(u) = \rho_s + \frac{\dot{\rho_s}}{\rho_s} + \cdots$. The other important variable is the energy. The total energy of the solution is given by the holographic Brown-York stress tensor (\ref{Etimelike}), in the limit as $r \rightarrow \infty$. Here there is a divergence which must be canceled by a counterterm. The resulting total energy turns out to be given by
\begin{align}
 E = -\frac{1}{8\pi G_N} \lim_{\rho \rightarrow \infty} \rho \left(n^A \nabla_A \phi - \phi \right).   \label{Eholo}
\end{align}
For the dilaton $\phi(u,\rho)$, the $\rho \rho$ component of the metric field equation (\ref{metricfieldeqn}) implies $\phi(u,\rho) = \rho \phi_r + A(u)$. Inserting this form into the $u\rho$ component yields $A(u) = \phi'_r$. With this form, using (\ref{Eholo}) one finds that the total energy is proportional to $-\phi''_r + \rho_s^2 \phi_r$. The remaining $uu$ component of the Einstein equation is just the conservation equation $dE/du=0$. If we re-define $u$ such that $\phi_r$ is a constant, then we have the equation $d\rho_s/du= 0$.

The horizon action (\ref{action}) should encode information about the infrared degrees of freedom in the dual finite temperature theory. Following \cite{Nickel:2010pr}, one can think about this action as arising from integrating out UV degrees of freedom in the framework of the Wilsonian renormalization group. In holography one integrates out the bulk degrees of freedom and is left with a low energy theory consisting of Goldstone bosons (the reparametrization mode) and near horizon variables ($\phi_H$). 

Suppose we consider the gravitational action to have contributions from both the boundary and the horizon. This takes the form
\begin{align}
I_{tot} = \frac{1}{16\pi G_N} \left[ \int d^2x \sqrt{g} \left( \phi (R+ 2) + 2 \int du \phi_b K + 2 \int dt \phi_H \kappa \right) \right].
\end{align}
Imposing the dilaton equation of motion, we arrive at the following total on-shell action, which is the sum of (\ref{action}) and (\ref{boundaryaction})
\begin{align}
I_{bdry, tot} = \frac{1}{8\pi G_N}  \int du \left[ \phi_H(u) \left(\frac{\tau''}{\tau'} + \kappa_0 \tau' \right) - \phi_r(u) \left( \frac{\tau'''}{\tau'}- \frac{3}{2} \frac{\tau''^2}{\tau'^2} - \frac{\kappa_0^2}{2} \tau'^2 \right) \right].  \label{totalaction}
\end{align}
In writing this action we have identified the horizon time with the boundary time. The horizon terms proportional to $\phi_H$ are the lowest order terms in an expansion in derivatives of the reparametrization mode. As expected, these terms characterize the infrared regime, while the terms from the Schwarzian action at the boundary describe the UV\footnote{A similar statement about the derivative expansion appears to be valid also in the case where $\kappa_0 = 0$ in (\ref{totalaction}), which corresponds to a boundary at the $z = \infty$ (extremal) horizon of the Poincare patch of $AdS_2$ (\ref{Poincare}). Here the temperature is identically zero though}. Note that both \cite{Maldacena:2016upp, Jensen:2016pah} added by hand to the Schwarzian action the term proportional to $\tau'$ in order to capture the extremal entropy of the $AdS_2$ black hole horizon. In the case where $\phi_H$ is a constant and the horizon is in identically in equilibrium, both terms from the horizon are topological.

Following the same procedure as in Section \ref{AdSBoundary},  it is easy to show that (\ref{totalaction}) can be expressed in hydrodynamical sigma model form. The result is
\begin{align}
I_{eff} = \frac{1}{8\pi G_N} d\tau \sqrt{-h} \left[  \phi_H \left(\frac{\dot{T}}{T} + 2\pi T \right) + \phi_r \left(\frac{3}{2} \frac{\dot{T}^2}{T^2} - \frac{\ddot{T}}{T} + 2 \pi^2 T^2 \right) \right].   \label{hydroaction}
\end{align}
Again, the horizon terms are the lowest order terms in an expansion in temperature and its derivatives. There is a freedom to add a constant ground state energy $E_0$, which will not affect the equations of motion. If we go to the equilibrium frame where $\tau = u$ and rotate to Euclidean signature, one finds the expected partition function and Wald entropy proportional to $\phi^{(0)}_H$. 

The total action is composed of the reparametrization mode acting as the field, plus two external couplings in the UV and IR. If we vary the action with respect to $\tau$ we find the sum of the horizon constraint equation and the field equation capturing the boundary dynamics,
\begin{align}
\left(\frac{\phi'_H}{\tau'}\right)' - \kappa_0 \phi'_H - \left[\frac{1}{\tau'} \left(\frac{(\tau' \phi_r)'}{\tau'}\right)'  \right]' - \kappa_0^2 (\phi_r \tau')' = 0
\end{align}
This equation implies there is a relationship between the UV variables and IR horizon value $\phi_H$. If we consider the case where $\tau(u) = u$ this reduces to
\begin{align}
\phi''_H - \kappa_0 \phi'_H - \phi'''_r - \kappa_0^2 \phi'_r = 0
\end{align}
From this equation we find that there is a relationship between the UV and IR values
\begin{align}
\phi_H = \kappa_0 \phi_r + \phi'_r. \label{UVIR}
\end{align}
This relation also follows from the radial evolution parts of the metric field equation we discussed above, where $\phi(u,\rho)=\rho \phi_r + \phi'_r$. In writing (\ref{totalaction}) we have consistently coupled the UV and IR degrees of freedom. With the condition (\ref{UVIR}) imposed, the boundary constraint equation for $\phi_r$ is encoded in the horizon constraint equation. Note that if we were to choose $\phi_r = 1$ then $\phi_H = \rho_h(u)$. When $\tau=u$, $\phi_H = \rho_0$, which is equilibrium.

Finally, we can also consider the symmetries and conserved charges of the UV and IR parts of the action. The UV part of the action is invariant under the three-parameter transformation
\begin{align}
\tau \rightarrow \tau + G + F e^{-\kappa_0 \tau} + H e^{\kappa_0 \tau}   \label{UVtransform}
\end{align}
which can be thought of as a generalization of the $SL(2,R)$ symmetry in the vacuum $AdS_2$ case to the thermal state dual to the $AdS_2$ black hole.This symmetry can be inferred from the solution to the constraint equation for $\phi_r$ in the case where $\tau = u$, which is $\phi'''_r - \kappa_0^2 \phi'_r = 0$. This yields
\begin{align}
\phi_r = c_0 + c_1 e^{\kappa_0 u} + c_2 e^{-\kappa_0 u}.   \label{phisol}
\end{align}
The three parameter transformation is the subset of infinitesimal diffeomorphisms that map one solution for $\phi_r$ into another.

The Noether charges were found in \cite{Maldacena:2016upp}, where they were labeled as $Q^0$ and $Q^{\pm}$ and interpreted in terms of symmetries of the thermofield double state of the extended $AdS_2$ black hole geometry.  When we consider the infrared terms in the action, we see that the symmetry reduces to just the two parameter family of $(F,G)$ described earlier, which correspond to $Q^0$ and $Q^{-}$. This happens because we are considering asymptotic symmetries near the future horizon in one wedge of the geometry.  If we were to consider the other wedge or the past horizon (via outgoing Eddington-Finkelstein coordinates) the relevant symmetries would instead involve the other two parameter family $(G,H)$. For the charges, one finds, for example
\begin{align}
Q_G =&\frac{1}{8\pi G} \left( \left(\frac{\phi_r}{\tau'}\right)'' - \phi_r \left(\frac{\tau'''}{\tau'^2} - 3 \frac{\tau''^2}{\tau'^3} + \kappa_0^2 \tau' \right)  \right) \\
Q_F =& \frac{1}{8\pi G} e^{-\kappa_0 \tau} \left(-\phi_r \frac{\tau'''}{\tau'^2} + 3 \phi_r \frac{\tau''^2}{\tau'^3}  + 3 \phi_r \kappa_0 \frac{\tau''}{\tau'} + 3 \left(\frac{\phi_r \tau''}{\tau'^2}\right)' + \kappa_0 \tau' \left(\frac{\phi_r}{\tau'}\right)' + \left(\frac{\phi_r}{\tau'}\right)'' \right) .
\end{align}
$Q_G$ is the total ADM energy. If we consider $\tau=u$, this reduces to $T_0 s_0$ in terms of horizon data. For $Q_F$, we can take $\tau=u$ as well. Then for the above solution (\ref{phisol}),
\begin{align}
Q_F =& 2 \kappa_0^2 c_1.
\end{align}
The charge is only sourced by the exponentially growing mode, which also mirrors the behavior we found earlier in Section III at the horizon. Maldacena, Stanford, and Yang showed the exponentially growing mode is associated with a ghost mode, which they argued can be dealt with by treating (\ref{UVtransform}) as a gauge symmetry of the quantum state. Taking into account the left wedge in the extended geometry, the charges vanish and the doubled state is invariant. However, for the wedge geometry associated with the mixed thermal state the charges do seem to have physical consequences. The existence of these modes was crucial to the exponentially growing behavior of the out of time ordered correlators, e.g. $\sim e^{\kappa_0 u}$ for Lyapunov exponent $\kappa_0$, used diagnose chaotic behavior in the dual (SYK model) theory. Here we can also see this type of behavior locally at the horizon using our infrared action, which implies that, for example, $E_{memb}$ is an exponentially increasing variable.

\section{Three-dimensional Gravity case}

Now that we have investigated the two-dimension dilaton gravity case, it is interesting to see whether these results generalize into the usual GR setting. We first consider the case of horizons in three-dimensional gravity. In this case the horizon supertranslations generalize to the transformation $\bar{v} = \alpha(v,x)$ which acts on the metric (\ref{nullgaussian}). This is no longer just a time reparametrization on the horizon. Note that we also have the superrotation freedom, which is a spatial diffeomorphism $\bar{x} = R(x)$. If we contract the Einstein field equation with two null normal vectors, we find the Raychaudhuri equation
\begin{align}
R_{AB} \ell^A \ell^B = \ell^A \nabla_A \theta - \kappa \theta + \theta^2 = 0.
\end{align}
Here there is no shear since the horizon cross-section is one-dimensional. Re-expressing this equation in terms of the derivative of the cross-sectional metric $\sqrt{\gamma}$ using the definition of the horizon expansion,
\begin{align}
\theta = \frac{{\cal L}_\ell \sqrt{\gamma}}{\sqrt{\gamma}},
\end{align}
yields
\begin{align}
\ell^A \nabla_A (\ell^B \nabla_B \sqrt{\gamma}) - \kappa \ell^A \nabla_A \sqrt{\gamma} = 0 \label{horizonconstraint3D}
\end{align}
This has the same form as the two-dimensional horizon constraint equation, via the replacement $\phi_H \rightarrow \sqrt{\gamma}$. Indeed, $\phi_H$ is proportional to the entropy for a two-dimensional black hole, while in higher dimensions the Bekenstein-Hawking entropy density is proportional to $\sqrt{\gamma}$.

To solve (\ref{horizonconstraint3D}) we again take $\ell^A \nabla_A = \partial_{\bar{v}}$. The solution to the constraint equation is then (for the simplest case of $\kappa = \kappa_0$)
\begin{align}
\sqrt{\gamma}(\bar{v},x) = c_0(x) + \frac{c_1(x)}{\kappa_0} e^{\kappa_0 \bar{v}},
\end{align}
which can be re-expressed as
\begin{align}
\sqrt{\gamma}(v,x) = c_0(x) + \frac{c_1(x)}{\kappa_0} e^{\kappa_0 \alpha(v,x)}.
\end{align}
This is the generalization of (\ref{dilatonfixing2}).

In this case there is no explicit breaking of the horizon asymptotic symmetries, but there is spontaneous breaking associated with the choice of state. As before, the set of diffeomorphisms that map one solution to another, or equivalently act as symmetries of the constraint equation, are associated with subset of supertranslations that preserve the surface gravity. At the infinitesimal level, the subset $v \rightarrow v + \epsilon(v,x)$  that solve the no change condition in (\ref{kappachange}) has the form
\begin{align}
\epsilon(v,x) = F(x) e^{-\kappa_0 v} + G(x).
\end{align}
The nature of the supertranslation $G(x)$ was discussed in \cite{Eling:2016xlx}, we saw that this leads to a shift in the one-form $\Omega_x$ via
\begin{align}
\Omega_x \rightarrow \Omega_x  -  \kappa_0 \partial_x G.
\end{align}
Using the membrane paradigm to identify $\Omega_i$ with the momentum $P_i$ of a non-relativistic horizon field theory system, we argued the spontaneous breaking of the supertranslations is a spontaneous breaking of a particle number $U(1)$ type of symmetry.

As in the two-dimensional case, one can construct an effective action whose variation yields the horizon constraint equation. In the case of a null boundary in GR it was argued, see e.g. \cite{Parattu:2015gga} that the proper form of the Gibbons-Hawking term is in general dimension $D$
\begin{align}
I_{null} =  \frac{1}{8\pi G_N}  \int dv d^{D-2} x \sqrt{\gamma} (\kappa + \theta).
\end{align}
Neglecting the $\theta$ term as a total derivative, we propose that in three-dimensions the relevant action is
\begin{align}
I_{eff} =  \frac{1}{8\pi G_N}  \int dv dx \sqrt{\gamma}  \left(\frac{\partial^2_v \alpha}{\partial_v \alpha}  + \kappa_0 \partial_v \alpha \right). \label{action2}
\end{align}
The equation of motion for $\alpha$ yields
\begin{align}
\partial^2_t \sqrt{\gamma} - \left(\frac{\partial^2_v \alpha}{\partial_v \alpha} + \kappa_0 \partial_v \alpha \right) \partial_v \sqrt{\gamma} = 0
\end{align}
which is the horizon constraint equation. As in the two-dimensional case, we can also write down an action for the extremal case where $\kappa_0=0$, which is just
\begin{align}
I_{eff} =  \frac{1}{8\pi G_N}  \int d\lambda dx \sqrt{\gamma}  \left(\frac{\partial^2_\lambda \alpha}{\partial_\lambda \alpha} \right). \label{extremalaction2}
\end{align}

The finite temperature action (\ref{action2})  is invariant under the global transformation $\alpha \rightarrow \alpha + G(x) + F(x) e^{-\kappa_0 \alpha}$. The resulting Noether charge densities are
\begin{align}
Q_{den} = \frac{1}{8\pi G_N} \left[ G(x) \left(\kappa_0 \sqrt{\gamma} - \frac{\partial_v \sqrt{\gamma}}{\partial_v \alpha} \right) - F(x) \frac{e^{-\kappa_0 \alpha}}{\partial_v \alpha} (\partial_v \sqrt{\gamma})\right].
\end{align}
Here we have an infinite set of charges. For  $\partial_v \alpha = 1$, these charges agree with those found by different methods in \cite{Donnay:2016ejv} . In the case where $G(x)$ is a constant, and we are in equilibrium, we find as before that the Noether charge density is proportional to entropy density. The higher harmonics of $G(x)$ (thinking of the horizon with spherical/circular symmetry) will yield a vanishing contribution. When $F(x)$ is a constant we can argue that the charge is associated with the membrane paradigm energy density \cite{Price:1986yy}, which is exponentially growing,
\begin{align}
\rho _{memb} = -\frac{1}{8 \pi G_N} \theta.
\end{align}
Integrating to find the total energy yields
\begin{align}
E_{memb} =  -\frac{1}{8 \pi G_N}  \int dx  \sqrt{\gamma} \theta = -\frac{1}{8 \pi G_N}  \int dx {\cal L}_\ell \sqrt{\gamma}.
\end{align}
We see that in this case ${\cal L}_\ell \sqrt{\gamma} $ plays a similar role to that of $\phi_H'$. In this case the quantity $E_{memb} e^{-\kappa_0 \alpha}$ is conserved via the horizon constraint equation. For the Noether charges, we find in general
\begin{align}
Q_{den} =  G(x) \left(T_0 s + \rho_{memb} \right) +  F(x) \rho_{memb} e^{-\kappa_0 \alpha} .
\end{align}
Since $G(x)$ is associated with the breaking of a $U(1)$ symmetry in the horizon system, we conjecture that the contribution from $G(x)$ in the out of equilibrium case is related to the particle number of the non-relativistic system.

In the extremal case, the action (\ref{extremalaction2}) is invariant under generalized global affine transformations $\alpha \rightarrow \alpha + G(x) + F(x) \alpha$. The resulting Noether charges can be expressed as
\begin{align}
Q^{ext}_{den} = G(x) \rho_{memb} + F(x) (s + \rho_{memb} \alpha). 
\end{align}

We can attempt to express the action (\ref{action2}) in  hydrodynamical form, following the two-dimensional dilaton example. In this case we have two-dimensional physical and reference manifolds. If we assume that we are implicitly in a reference frame where the vector field $\beta^a = \beta^\alpha = \beta_0$ (purely in the timelike direction), then the one-dimensional hydrodynamical action generalizes naturally
\begin{align}
I_{eff} = \frac{1}{8\pi G_N} \int d\alpha dx \sqrt{-h} \sqrt{\gamma} \left(\frac{\dot{T}}{T} + 2\pi T - E_0 \right).  \label{hydroaction2}
\end{align}
However, this doesn't seem to capture the dynamics entirely since it is possible to have spatial gradients in the system, via terms like $P^b_a \partial_b T$, where $P^{ab} = h^{ab} + u^a u^b$ is the transverse projector.

Finally, one can also add matter to the system just as in the two-dimensional case. We expect matter fields to also carry the conserved Noether charges. For scalar matter one can write down an effective action
\begin{align}
I_{eff} = \int dv dx ~\psi_H \left(\frac{\partial^2_v \alpha}{\partial_v \alpha}  + \kappa_0 \partial_v \alpha \right)
\end{align}
which captures their contribution to the Raychaudhuri equation.

\section{$D>3$ Gravity?}

In this case, the story with supertranslations is a simple generalization of the $D=3$ case. However, here the horizon constraint equation/Raychaudhuri equation has the form
\begin{align}
R_{AB} \ell^A \ell^B = \frac{d\theta}{dt} - \kappa \theta + \frac{1}{D-2} \theta^2 + \sigma_{ab} \sigma^{ab} = 0.
\end{align}
In this case we can no longer re-express this equation in the form of (\ref{horizonconstraint3D}) due to the different factor in front of the $\theta^2$ term. In higher dimensional GR there are dynamical degrees of freedom. In particular, there is also a shear term, which can be thought of as encoding the contribution from the flux of gravitational waves across the horizon. This acts like a dissipation term in the horizon system. 

Neglecting the shear squared term for the moment, we find
\begin{align}
\partial^2_{\bar{t}} \sqrt{\gamma} - \kappa \partial_{\bar{t}} \sqrt{\gamma} + A \frac{(\partial_{\bar{t}} \sqrt{\gamma})^2}{\sqrt{\gamma}} = 0,
\end{align}
where $A = \frac{D-3}{D-2}$. The solution is
\begin{align}
\sqrt{\gamma} = \left((A+1) \left(\frac{F(x^i)}{\kappa} e^{\kappa \alpha} + G(x^i)\right)\right)^{\frac{1}{A+1}}.
\end{align}
It is not clear whether one can construct an action whose variation will give back this equation of motion and solution. For small perturbations $\epsilon$ one can write down, for example,
\begin{align}
I_{eff} \sim \int dv d^{D-2}x~ \sqrt{\gamma} \left(\partial_v^2 \epsilon + \kappa_0 \partial_v \epsilon - A \theta \partial_v \epsilon  - A \epsilon \partial_v \theta \right).
\end{align}
On the other hand, linearized perturbations of the horizon are governed by (\ref{horizonconstraint3D}), where one drops the $\theta^2$ term as higher order. One could also absorb the shear squared term into any generalized matter-energy flux. Another alternative is to work at lowest orders in a (hydrodynamic) expansion in time derivatives. In this setting the effective action would generate the ideal entropy conservation equation $\partial_v \sqrt{\gamma} = 0$. Higher order dissipative corrections to this equation, which involve the $\theta^2$ and $\sigma^2$ terms, would not be captured. Thus, we conclude that the horizon effective action can likely only describe the horizon dynamics in the case where the metric degrees of freedom are only slightly perturbed, or at lowest orders in a hydrodynamic expansion.

\section{Discussion}

In this paper we constructed an effective action for the Goldstone modes associated with the spontaneous breaking of horizon supertranslations. The variation of this action produces the horizon constraint/Raychaudhuri equation in two and three-dimensional gravity theories, but only approximately in higher dimensions. We found that this action can be expressed in a hydrodynamical sigma model form and, in the context of holography, it captures the low-energy degrees of freedom near the horizon. 

One may wonder if a similar type of effective action exists for the spontaneous breaking of horizon superrotations in three and higher dimensions. Under a superrotation, the horizon metric $\gamma_{ij} \rightarrow \gamma_{ij} + {\cal L}_R \gamma_{ij}$, which implies that $\sqrt{\gamma} \rightarrow \sqrt{\gamma} + \partial_i R^i$. This means that for a state with a given area, there is a spontaneous breaking of the superrotations down to area-preserving (spatial) diffeomorphisms. The resulting effective action is therefore invariant under these symmetries and appears to be closely related to the effective actions for holographic ideal fluids constructed in \cite{Crossley:2015tka,deBoer:2015ija}. 

Finally, black holes in two and three dimensions are important toy models for understanding issues such as the information paradox and the microstate origin of the Bekenstein-Hawking entropy. In the future it would be interesting to understand whether these soft mode ``hairs" on non-stationary horizons play any role in the resolution of these problems.

\section*{Acknowledgements}
I would like to thank Yaron Oz for valuable discussions.
This research was supported by the European Research Council
under the European Union's Seventh Framework Programme (ERC Grant
agreement 307955).

\end{document}